\documentclass[11pt,twoside]{article}
\usepackage{newpasp,epsf}
\def\edcomment#1{\iffalse\marginpar{\raggedright\sl#1\/}\else\relax\fi}
\reversemarginpar

\begin{document}
\title{Estimating jet power in proton blazar models}
\author{R.J. Protheroe} \affil{Department of Physics and
Mathematical Physics, The University of Adelaide, Adelaide, SA
5005, Australia}

\author{A. M\"ucke}
\affil{Universit\'e de Montr\'eal, D\'epartement de Physique,
Montr\'eal, H3C 3J7, Canada}

\begin{abstract}
We discuss the various contributions to the jet luminosity in
proton blazar models of active galactic nuclei and describe a
method of estimating the jet luminosity from the observed
spectral energy distribution (SED) and the fitted model
parameters.  We apply this to a synchrotron proton blazar (SPB)
model for Markarian 501.
\end{abstract}

\section{Introduction}

The mechanical luminosity of an AGN jet of cross-sectional area
$A$ with Lorentz factor $\Gamma=1/\sqrt{1-\beta^2}$, containing
jet-frame (primed variables) matter density density $\rho'$,
energy density $u'$ and pressure $p'$ is given by
\begin{eqnarray*}
L_{\rm jet,\, mech} &=& \Gamma^2 \beta c
A[\rho'c^2(\Gamma-1)/\Gamma + u'+p'].
\end{eqnarray*}
(see e.g. Leahy 1991).  To calculate the total jet luminosity
$L_{\rm{jet}}$, measured in the rest frame of the galaxy, we
adapt the formulae of Bicknell (1994) and Bicknell \& Dopita
(1997), given for the synchrotron self-Compton model, to apply
for the case of proton blazar models.  In this paper we shall
apply the formula adapted for proton blazar models to a
synchrotron proton blazar (SPB) model for Markarian 501 described
by M\"ucke \& Protheroe (2000).

\section{Proton blazar models}

In proton blazar models, the high energy part of the SED is due
to interaction of protons accelerated along with electrons in the
AGN jet.  The interactions are pion photoproduction on either low
energy photons of the low energy part of the SED produced as
synchrotron radiation by electrons in the jet (e.g. Mannheim
1993), or on direct or scattered UV bump radiation from an
accretion disk (e.g. Protheroe 1997), and direct synchrotron
emission by protons, muons and charged pions (e.g. M\"ucke \&
Protheroe 2000). To accelerate protons to sufficiently high
energies that they can produce the high energy part of the SED, a
large magnetic field is required.

Proton blazars would contain relativistic plasma of electrons and
protons, and a non-negligible magnetic field.  For a
$n'_e({\gamma_e}')={n_e}_0{{\gamma_e}'}^{-2}$ power-law
(${\gamma_e}'_1 < {\gamma_e}' < {\gamma_e}'_2$) for the jet-frame
number density of electrons, the total number density of
(relativistic) electrons is $n'_e = {n_e}_0/{\gamma_e}'_1$, and
the energy density of (relativistic) electrons is $3p'_e=u'_e =
n'_e m_ec^2 \langle {\gamma_e}' \rangle = {n_e}_0 m_ec^2
\ln({\gamma_e}'_2/{\gamma_e}'_1)$.  Similarly, for a
$n'_p({\gamma_p}')={n_p}_0{{\gamma_p}'}^{-2}$ power-law
(${\gamma_p}'_1 < {\gamma_p}' < {\gamma_p}'_2$) for the jet-frame
number density of protons, the total number density of
(relativistic) protons is $n'_p = {n_p}_0/{\gamma_p}'_1$, and the
energy density of (relativistic) protons is $3p'_p=u'_p = n'_p
m_pc^2 \langle {\gamma_p}' \rangle = {n_p}_0 m_pc^2
\ln({\gamma_p}'_2/{\gamma_p}'_1)$, giving
\begin{eqnarray*}
n'_e m_e c^2 \approx {3p'_e \over {\gamma_e}'_1
\ln({\gamma_e}'_2/{\gamma_e}'_1)}, \;\;\; 
n'_p m_p c^2 \approx {3p'_p \over {\gamma_p}'_1
\ln({\gamma_p}'_2/{\gamma_p}'_1)}.
\end{eqnarray*}
Assuming that the number of relativistic electrons will be
greater than the number of relativistic protons, and applying
charge conservation, i.e. the number of `cold' (non-relativistic)
protons equals the number of `hot' (relativistic) electrons minus
the number of hot protons, one obtains
\begin{eqnarray*}
\lefteqn{L_{\rm jet,\, mech} =} \\ 
& &=\Gamma^2 \beta c
A\left[\left({m_p \over m_e} {3p'_e \over {\gamma_e}'_1
\ln({\gamma_e}'_2/{\gamma_e}'_1)} - {3p'_p \over {\gamma_p}'_1
\ln({\gamma_p}'_2/{\gamma_p}'_1)}\right){(\Gamma-1)\over \Gamma}
+ 4p_e' + 4p_p' + 4p_B' \right]\\ &  &= 4p_p'\Gamma^2 \beta c
A\left[\chi_p {(\Gamma-1)\over\Gamma} + 1 + {p_B'\over p_p'}+
{p_e'\over p_p'} \right]
\end{eqnarray*}
since for a tangled magnetic field $u_B'=3p_B'$, and
where
\begin{eqnarray*}
\chi_p = {3\over4}\left({m_p \over m_e}{p'_e \over p'_p} {1 \over
{\gamma_e}'_1 \ln({\gamma_e}'_2/{\gamma_e}'_1)} - {1 \over
{\gamma_p}'_1 \ln({\gamma_p}'_2/{\gamma_p}'_1)}\right).
\end{eqnarray*}
We now consider how to apply these formulae to proton blazar
models.  First, consider the radiation efficiency of protons
which depends on their effective energy loss rates $r'$ for
synchrotron radiation (syn), photoproduction leading to
electromagnetic radiation (EM) or anything (any), and adiabatic
losses (adiab),
\begin{eqnarray*}
\zeta'_p(\gamma_p') = { r_{\rm syn, p}'(\gamma_p') + 
r_{\rm p\gamma\to EM, p}'(\gamma_p') \over r_{\rm syn, p}'(\gamma_p') + 
r_{\rm p\gamma\to any, p}'(\gamma_p') + r_{\rm adaib}'}.
\end{eqnarray*}
Averaging over the input {\em energy} spectrum gives the total
radiation efficiency
\begin{eqnarray*}
\zeta_p = {\int \zeta'_p(\gamma_p') {\gamma_p}'{{\gamma_p}'}^{-2}
d {\gamma_p}' \over \int {\gamma_p}'{{\gamma_p}'}^{-2} d
{\gamma_p}'}.
\end{eqnarray*}
For electrons, we assume the radiation efficiency  to be $\zeta_e=1$. 

If we assume that the low energy part of the observed SED is due
to electrons, and the high energy part is due to protons, then we
may infer values of $p_e'$ and $p_p'$ directly from the observed
bolometric flux in the two parts of the SED for an assumed
Doppler parameter and emission region size. The bolometric
luminosities for the two parts of the SED (in any frame since
they are Lorentz invariant) are then $L^{\rm low}_{\rm bol} =
\zeta_e\Gamma^2 \beta c A 4p_e'$ and $L^{\rm high}_{\rm bol} =
\zeta_p \Gamma^2 \beta c A 4p_p'$.  These bolometric luminosities
are related to the observed bolometric fluxes (corrected for
relativistic beaming) from the two parts of the SED by $L^{\rm
low}_{\rm bol} = 4 \pi d_L^2 S^{\rm low}_{\rm obs}/D^2$ and
$L^{\rm high}_{\rm bol} = 4 \pi d_L^2 S^{\rm high}_{\rm
obs}/D^2$ with $D=[\Gamma(1 - \beta \cos \theta)]^{-1}$ the
Doppler factor and $d_L$ the source's luminosity distance.  Hence,
we obtain
\begin{eqnarray*}
L_{\rm jet,\, mech} &=& {L^{\rm high}_{\rm obs} \over D^2
 \zeta_p} \left[\chi_p {(\Gamma-1)\over\Gamma} + 1 + {p_B'\over
 p_p'}+ {\zeta_p S^{\rm low}_{\rm obs} \over \zeta_e S^{\rm
 high}_{\rm obs}} \right]
\end{eqnarray*}
where $p_B'=[{B'}^2/(2\mu_0)]/3$, and
\begin{eqnarray*}
p_p' = {L^{\rm high}_{\rm obs} \over 4 D^2 \zeta_p \Gamma^2
\beta c A}, \;\;\;\; \chi_p = {3\over4}\left({m_p \over
m_e}{\zeta_p S^{\rm low}_{\rm obs} \over \zeta_e S^{\rm
high}_{\rm obs}}{1 \over {\gamma_e}'_1
\ln({\gamma_e}'_2/{\gamma_e}'_1)} - {1 \over {\gamma_p}'_1
\ln({\gamma_p}'_2/{\gamma_p}'_1)}\right).
\end{eqnarray*}

\section{Application to SPB model for Markarian 501}

We use these formula to estimate the total jet power of Mrk 501
during its 1997 flare to be $\sim 10^{46}$~erg/ s, and find the
contributions to the total jet power of cold protons, magnetic
field, and accelerated electrons, relative to that of accelerated
protons.  Fig.~1 shows the dependence of the total jet luminosity
on the Doppler factor $D$ for a fixed variability time scale
$t_{\rm{var}} = 12$~hours, and with $B$, ${n_p}'$ and a target
photon density appropriate to Mrk~501 during flaring.  Clearly
visible is the fact that in hadronic models the proton kinetic
energy and the poynting flux dominate the total jet luminosity,
while the electron kinetic energy is only of minor importance. At
high Doppler factors the emission region becomes so large that
one needs only relatively small magnetic fields and proton
densities to fit the observations. In addition, adiabatic losses
become small resulting in a decrease of the required kinetic
proton luminosity.  For example, $B\approx 5$~G and $n_p' \approx
10^{-2}$~cm$^{-3}$ are sufficient to fit the Mrk~501 flare for
$D=50$, while for $D=8$ magnetic fields of over 30~G and proton
densities of $n_p' \approx 10^{4}$~cm$^{-3}$ are needed. The
total jet luminosity exhibits a minimum of $\sim 10^{46}$erg/s at
$D\approx 12$.

\begin{figure}[ht] 
\plotfiddle{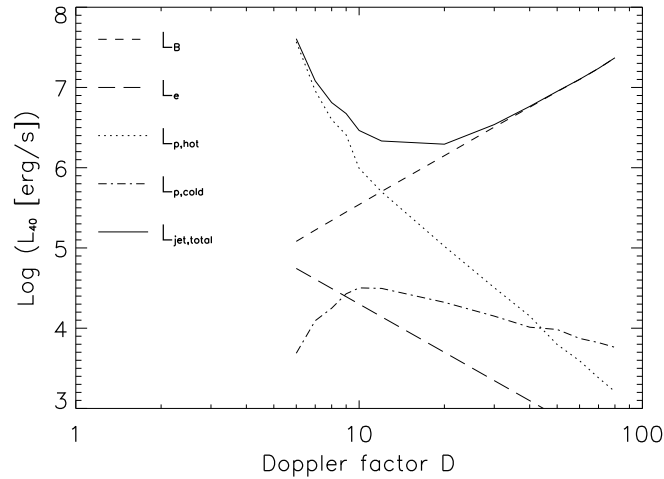}{60mm}{0}{56}{56}{-185}{-208}
\caption{Dependence of total jet luminosity $L_{\rm{jet}}$ and
its contributions ($L_B$ -- magnetic field, $L_{\rm{p,hot}}$ and
$L_{\rm{p,cold}}$ -- hot and cold protons, $L_{\rm{e}}$ --
electrons) on $D$ for model parameters which reasonably fit the
Mrk~501 1997 flare SED (see M\"ucke \& Protheroe 2000 for
details). $L_{40} = L/10^{40}$~erg/s.  }
\label{fig1}
\end{figure}

In the framework of the jet--disk symbiosis (e.g. Falcke \&
Biermann 1995), the jet luminosity should not exceed the total
accretion power $Q_{\rm{accr}}$ for the equilibrium state.
Accretion theory relates the disk luminosity to the accretion
power. Page \& Thorne (1974) give $L_{\rm{disk}} \approx
(0.05-0.3) Q_{\rm{accr}}$.  Disk luminosities for `typical'
radio-loud AGN lie in the range $L_{\rm{disk}} \approx
10^{44}-10^{48}$~erg/s with BL~Lac objects tending to the lower
end on average. Specifically, for Mrk~501 there are no emission
line measurements available, and this complicates the evaluation
of its disk luminosity. However, any observed
UV-emission in the flaring stage may put an upper limit on it.
Historical data give $L_{\rm{disk}} \approx
10^{43}-10^{44}$~erg/s (Mufson et al 1984, Pian et al 1998), and
we obtain for the accretion power, $Q_{\rm{accr}} \approx
(3-200)\times 10^{43}$~erg/s, at least a factor 5 below the value
necessary to comply with the constraint of the disk--jet
symbiosis. Note, however, that the estimate of $Q_{\rm{accr}}$ is
based on archival non-flaring data from Mrk~501, and we could
speculate that either the disk has pushed more energy into the
jet during TeV-flaring, or that the flaring stage can not be
considered as a steady state.  Also, accretion theory might
predict larger conversion efficiencies of the accretion power
into disk radiation than actually might occur in BL~Lac objects.

\end{document}